\documentclass[12pt]{article}

\usepackage{epsfig}
\usepackage{cite}

\usepackage{amsmath}
\usepackage{amssymb}
\usepackage{euscript}

\newcommand{\Pcal}{{\cal P}}
\newcommand{\Peu}{\EuScript{P}}

\begin{document}

\begin{titlepage}


\begin{flushright}
\bf IFJPAN-IV-2007-4\\
    CERN-PH-TH/2007-064
\end{flushright}

\vspace{1mm}
\begin{center}
  {\LARGE\bf%
    Markovian\vspace{1.9mm} Monte Carlo solutions\\
    of the one-loop CCFM equations$^{\star}$
}
\end{center}
\vspace{2mm}

\begin{center}
{\large\bf K.~Golec-Biernat$^{ac}$, S.~Jadach$^{ad}$, W.~P\l{}aczek$^b$,\\
P.~Stephens$^{a}$
{\rm and} \large\bf M.~Skrzypek$^{ad}$}
\end{center}
\vspace{2mm}

\begin{center}
{\em $^a$Institute of Nuclear Physics IFJ-PAN,\\
  ul.\ Radzikowskiego 152, 31-342 Cracow, Poland.}\\ \vspace{2mm}
{\em $^b$Marian Smoluchowski Institute of Physics, Jagiellonian University,\\
   ul.\ Reymonta 4, 30-059 Cracow, Poland.}\\ \vspace{2mm}
{\em $^c$Institute of Physics, University of Rzeszow,\\
   ul.\ Rejtana 16A, 35-959 Rzeszow, Poland.}\\ \vspace{2mm}
{\em $^d$CERN, PH Department, CH-1211 Geneva 23, Switzerland.} 
\end{center}

\vspace{3mm}
\begin{abstract}
A systematic extension of the Monte Carlo (MC) algorithm,
that solves the DGLAP equation, into the so-called
the one-loop CCFM evolution is presented.
Modifications are related to a $z$-dependent coupling constant;
transverse momentum dependence is added
to the $x$-dependence of the parton distributions.
The presented Markovian algorithm for one-loop CCFM evolution is
the first step in extending it to other
more sophisticated schemes beyond DGLAP.
In particular, implementing the complete CCFM will be the next step.
The presently implemented one-loop CCFM option
will be a useful tool in testing the forthcoming MC solutions.
Numerical results of the new MC are confronted with 
other non-MC numerical solutions.
The agreement within the MC statistical error of $\sim 0.1\%$ is found.
Also, numerical results for $k^T$-dependent structure functions are presented.
\end{abstract}

\vspace{3mm}
\begin{center}
\em To be submitted to Acta Physica Polonica
\end{center}

\vspace{3mm}
\begin{flushleft}
\bf IFJPAN-IV-2007-4\\
    CERN-PH-TH/2007-064
\end{flushleft}

\vspace{5mm}
\footnoterule
\noindent
{\footnotesize
$^{\star}$The project is partly supported by the EU grant MTKD-CT-2004-510126,
  realized in the partnership with the CERN Physics Department and by the
  Polish Ministry of Science and Information Society Technologies
  grant No 620/E-77/6.PR UE/DIE 188/2005-2008.
}

\end{titlepage}

\section{Introduction}
The evolution equations (EVEQs) play a crucial role in the construction of
any parton shower Monte Carlo program simulating
the production of multiple gluons and quarks, in an approximate way,
in perturbative QCD.
In particular, EVEQs are in this context used to model the actual
development of the tree-like cascade of primary emitted partons from the
initial hadron. Such a cascade is then supplemented
with an appropriate hard scattering matrix element and with
the hadronization mechanism to form the complete parton shower MC code.

The basic, and by far the best analyzed, EVEQs in QCD are the DGLAP
equations \cite{DGLAP}; originally formulated in the leading order, later
extended to the next-to-leading order
\cite{Floratos:1977au,Floratos:1978ny,Curci:1980uw,Furmanski:1980cm}
and recently  
even to the next-to-next-to-leading order \cite{Moch:2004pa,Vogt:2004mw}. 
These equations perform
systematic resummation in terms of the variable $\log Q$, where $Q$ is
a hard scale, 
for not too small or too large values of the longitudinal momentum fraction $x$.

Another important EVEQ is the BFKL equation \cite{BFKL}.
Originally formulated in the leading
order, it has  been recently extended to the next-to-leading order~\cite{BFKL-nlo}.
Contrary to the DGLAP equations, it resums large $\log x$ 
terms, therefore, describing the small $x$ limit better than the DGLAP equations.
In recent years attempts have been made to improve the DGLAP evolution equations
in the small-$x$ region by incorporating some of the BFKL 
features~\cite{Altarelli:2005ni,Ciafaloni:2006yk}.
However, these attempts have lead to rather complicated evolution equations.

The third class of EVEQs in QCD are the CCFM equations \cite{CCFM}. 
They are formulated for the so called ``unintegrated'' parton
distributions, which depend on the
transverse momenta of partons $k^T$
in addition to $x$ and $Q$.
The main idea of the  CCFM approach  is to
correctly describe not only the large-$x$ region, where the summation
of $\log q$ dominates, but also the region of small $x$,
where the large logarithms
$\ln(1/x)$ are important. Thus, the CCFM equations
effectively interpolate between the DGLAP and BFKL equations.
One of the variants of the CCFM equations, called ``one-loop''
\cite{Marchesini:1991zy},  is of special interest because of its
simplicity and ease in extending to non-leading orders and
to non-gluonic evolutions.

The one-loop CCFM equation combines coherence
effects (angular ordering of gluonic emissions) at large $x$ with
the transverse momentum ordering at small $x$ \cite{Golec-Biernat:1997zs}.
In contrast, the ``all-loop" CCFM formulation extends the angular ordering 
to the small $x$ region.
In its original formulation the CCFM scheme has been formulated only for gluonic
cascades.
In the ``one-loop" approximation, as defined in ref.~\cite{Kwiecinski:2002},
it was extended to include also quarks into it.

Finally, let us mention yet another class of equations, the IREE ones
\cite{Ermolaev:2000sg}, which account for both
the logarithms of $Q$ and $x$ by means of constructing and
solving the so-called two-dimensional infrared evolution equations.

In a series of recent papers \cite{Jadach:2003bu, zinnowitz04,
raport04-06, raport04-07, Jadach:2005rd, Jadach:2006ku,
Golec-Biernat:2006xw, Jadach:2006yr, raport07-01}, 
a new family of Monte Carlo algorithms that solve the EVEQs in QCD
has been developed.
The novelty of these algorithms is in their ability to include an energy
constraint on top of the normal Markovian random walk-type
evolution. 
This is of the utmost importance in the case of the MC simulations of
narrow resonance production processes
initiated by initial state parton cascades.
In parallel to the ``constrained''
algorithms, we have also developed the standard ``unconstrained'' ones,
which are useful for tests and precise studies 
(for example, for fitting deep-inelastic structure function $F_2$).
In the very beginning of the above studies, the DGLAP evolution and some
of its variants were the main object of the interest.
However, from the previous discussion it is clear
that it would be useful and interesting to extend them also to other types
EVEQs, like those mentioned, having in mind the construction
of the parton shower Monte Carlo at a later stage.

In the presented paper we take a first step in that direction
and show how the Markovian ``unconstrained'' DGLAP algorithm can be extended
to the one-loop CCFM case. 
This extension involves two steps:
the modification of the coupling constant which becomes
$z$-dependent and the inclusion of transverse momenta into the evolution. 

In the subsequent papers further extensions of 
the Markovian algorithms will be presented, 
in particular to the all-loop CCFM scheme.
All technical aspect of the implementation of these
``constrained'' MC algorithms are discussed at length 
in ref.~\cite{Jadach:2007singleCMC}.
Summarizing, this work should be be treated as a warm-up exercise
for the forthcoming more 
complete and also more sophisticated analysis.

The paper is organized as follows. In Section 2 we recall some basic
formulae on the DGLAP equations and describe in detail problems to be
solved. Section 3 is devoted to the changes in the
algorithm related to the modification of the argument of the coupling
constant. Section 4 describes the extension of the algorithm to the
$k^T$-dependent evolution. In Section 5 some numerical results are
presented, and finally, Section 6 contains summary and outlook.

\section{Framework}

In this paper we will heavily rely on the notation and formulas from
\cite{Golec-Biernat:2006xw} where the reader will find more
details on the  notation and description of the framework which we use.
We start from  the DGLAP equations%
\footnote{This is eq.~(104) in ref.~\cite{Golec-Biernat:2006xw}.
   We shall provide explicit link to key formulas in this work in the following.}
\begin{equation}
  \label{eq:Evolu2}
  \partial_t\, xD_K(t,x)
   = \sum_J \int\limits_x^1 \frac{d z}{z}\; z\Peu_{KJ}(t,z)\; 
    \frac{x}{z} D_J\Big(t,\frac{x}{z} \Big)
\end{equation}
where $x$ the energy fraction of the hadron carried by the parton
of type $K=$ gluon, quark, antiquark.
The so-called evolution time is $t=\ln Q=\ln\mu$, with
$Q$ being large energy scale defined by the hard process probing
parton distribution function (PDF) $D_K(t,x)$.

The iterative solution solution of the above equations reads%
\footnote{See eq.~(105) in ref.~\cite{Golec-Biernat:2006xw}.}
\begin{equation}
  \label{eq:Iter6}
  \begin{split}
  xD_K&(t,x) = e^{-\Phi_K(t,t_0)} xD_K(t_0,x)
  +\sum_{n=1}^\infty \;
  \int\limits_0^1 dx_0\;
   \sum_{K_0,\ldots,K_{n-1}}
      \prod_{i=1}^n \bigg[ \int\limits_{t_0}^t dt_i\;
      \Theta(t_i-t_{i-1}) \int\limits_0^1 dz_i\bigg]
\\&~~~\times
      e^{-\Phi_K(t,t_n)}
      \prod_{i=1}^n 
          \bigg[ 
                 z_i\Peu_{K_i K_{i-1}}^\Theta (t_i,z_i) 
                 e^{-\Phi_{K_{i-1}}(t_i,t_{i-1})} \bigg]
      x_0 D_{K_0}(t_0,x_0) \delta\big(x- x_0\prod_{i=1}^n z_i \big),
  \end{split}
\end{equation}
where $K\equiv K_n$.
In order to turn the series of eq.~(\ref{eq:Iter6}) into the one-loop CCFM 
evolution we add two elements. First, the argument of coupling constant is made
$z$-dependent: $\alpha_s(t) \to \alpha_s(t+\log(1-z))$ and second,
the evolution has to include the transverse momenta of emitted
partons. As for the change of argument of $\alpha_s$ there are several
technical points to be clarified.

(i) We restrict ourselves to the LL
approximation only, so the kernels $\Peu_{K_i K_{i-1}} (t_i,z_i)$ have
the form%
\footnote{See eqs.~(50) and (A.2) in ref.~\cite{Golec-Biernat:2006xw}.}
\begin{equation}
\label{factor2}
\Peu_{KJ}(t,z)\,=\,2\/P_{KJ}(z,\alpha_s),
\end{equation}
\begin{equation}
P_{KJ}(z,\alpha_s) = \frac{\alpha_s^{(0)}(t+\ln(1-z))}{2\pi}\, P_{KJ}^{(0)}(z),
\end{equation}
where functions $P_{KJ}^{(0)}(z)$ are defined in eqs.\ [(A.5)].
The factor of 2 in eq.~(\ref{factor2}) is related to the definition of
the evolution time $t= \ln q$. 
The coupling constant $\alpha_s(t)$ is defined as follows%
\begin{equation}
  \label{eq:alfS1}
  \alpha^{(0)}_s(t)= \frac{4\pi}{ \beta_0 (2t-2\ln\Lambda_0)}.
\end{equation}

(ii) In the original papers 
\cite{Amati:1980ch, Brodsky:1982gc,Sterman:1986aj}
the shifted argument of $\alpha_s$ has been used only for the diagonal
kernels (Q$\to$Q or G$\to$G). We have, however, decided to apply the
same shift for all kernels, including the Q$\leftrightarrow$G ones
too.

(iii) In the presence of the $(1-z)$ factor the argument of the coupling
constant can become arbitrarily small. Therefore, to avoid the Landau
singularity we impose the following infrared (IR) cut-off
depending on the $q_0$ parameter:
\begin{equation}
\label{cutoff}
q_i(1-z_i)\geq q_0.
\end{equation}
It can be translated into $z_i\leq 1-q_0/q_i$ in space of $z$-variables.
The IR cut-off $\epsilon=q_0/q_i$ is not necessarily
infinitesimal anymore, for $q_i$ close to $q_0$.

(iv)
As a consequence of a finite IR cut-off on the real emissions, the momentum
sum rule 
\begin{equation}
\int dz \sum_J z \Peu^{}_{JK}(t,z)=0
\end{equation}
gets violated (by $\sim Q/q_0$ corrections).
In order to restore the sum rule we adjust
appropriately the virtual parts of the diagonal kernels
\begin{equation}
  \begin{split}
  \Peu_{KJ}(t,z)&=-\Peu^{\delta}_{KK}(t,q_0)\, \delta_{KJ}\,\delta(1-z)
                 +\Peu^{\Theta}_{KJ}(t,z),
\\
\Peu^{\Theta}_{KJ}(t,z)&=\Peu_{KJ}(t,z)\,\Theta(1-z-q_0e^{-t}),
  \end{split}
\end{equation}
where
\begin{equation}
\Peu^{\delta}_{KK} =\int\limits_0^1 dz\; z\sum_J \Peu^{\Theta}_{JK}(t,z).
\end{equation}
In the following sections we will describe in detail all of the above
technical points.

\section{The $z$-dependent strong coupling}
In this section we will describe the MC implementation of the
$z$-dependent coupling constant. Let us recall the probability
distributions of a single step forward in the Markov process.
The {Sudakov form factor} is defined in the usual way
\begin{align}
\Phi_K(t_i,t_{i-1};t_0) 
&= \int_{t_{i-1}}^{t_i}dt' \Peu_{KK}^\delta(t',t_0)
=
\int_{t_{i-1}}^{t_i}dt'\sum_J 
     \int_{0}^{1-\exp(t_0-t')}dz\; z\Peu^\Theta_{JK}(z)
\end{align}
and the probability of a { step forward in $z$} is%
\footnote{Following eq.~(69) of ref.~\cite{Golec-Biernat:2006xw}.}
\begin{equation}
\frac{d\omega (z_i,t_i,K_i;t_{i-1},K_{i-1})}{dt_idz_i}
   = z_i \Peu_{K_iK_{i-1}}^\Theta(z_i)  
              \theta_{1-e^{t_0-t_i}\geq z_i}
     e^{-\Phi_{K_{i-1}}(t_i,t_{i-1};t_0)} \theta_{t_i\geq t_{i-1}}.
\end{equation}
The probability of {a step forward in $t$} is given by the integral over $dz_i$ of
the above $d\omega$ 
\begin{equation}
\frac{d\omega (t_i;t_{i-1},K_{i-1})}{dt_i}
   = \theta_{t_i\geq t_{i-1}}\partial_{t_i}\Phi_{K_{i-1}} (t_i,t_{i-1};t_0)
     e^{-\Phi_{K_{i-1}}(t_i,t_{i-1};x_{i-1})}.
\end{equation}
In the above formulae we included also the new, finite cut-off
(\ref{cutoff}). 

\subsection{Singular part of the kernel}
\label{SubSec:step2}

Let us first calculate the Sudakov form factor 
only for the singular part of the kernels.
The approximate kernels are now 
\begin{equation}
\label{pcal}
{z\Peu}^{A\Theta}_{KK}(z,t;t_0)
= \frac{\alpha_s^{(0)}(t+\ln(1-z))}{\pi}
     \; \frac{\Theta(1-z-\epsilon(t))}{1-z}\;  A^{(0)}_{KK}\,,
            \;\;\;  K=G,q,\bar{q}\,,
\end{equation}
where $\epsilon(t)=\exp(t_0-t)$. The virtual components will be
discussed shortly.
The kernels of eq.~(\ref{pcal}) should appear in denominators of the
expression for the compensating weight $\bar{w}_P$.
In this particular case we are able to calculate the form factor
analytically in an exact way
\begin{equation}
\begin{split}
  {\Phi}^A_K(t_2,&t_1;\tau_0)
  =\int\limits_{t_1}^{t_2} dt'\;
     \int\limits_0^{1} dz\;
     \frac{\alpha_s^{(0)}(t'+\ln(1-z))}{\pi}\,
     A_{KK}^{(0)}\, \frac{\Theta(1-z-\epsilon(t'))}{1-z}
\\&=\frac{2}{\beta_0}
\int\limits_{t_1}^{t_2} dt'\;
     \int\limits^0_{t_0-t'} d \ln x\;
     \frac{1}{t'-\ln\Lambda_0 +\ln x}\,
     A_{KK}^{(0)}
\\&=\frac{2}{\beta_0}
    \int\limits_{t_1}^{t_2} dt'\;
    [\ln(t'-\ln\Lambda_0) -\ln(t_0-\ln\Lambda_0)]\, A_{KK}^{(0)}
    =\frac{2}{\beta_0}
    \int\limits_{t_1}^{t_2} dt'\;
    (\tau'-\tau_0)\, A_{KK}^{(0)}
\\&=\frac{2}{\beta_0} A_{KK}^{(0)}
    \int\limits_{\tau_1}^{\tau_2} d\tau'\;
    e^{\tau'}(\tau'-\tau_0)
   =\frac{2}{\beta_0}\, A_{KK}^{(0)}\;
    (\tau'-\tau_0-1)\,e^{\tau'}\bigg|_{\tau_1}^{\tau_2}
\\&= \frac{2}{\beta_0}\, A_{KK}^{(0)}\, \{\varrho(\tau_2)-\varrho(\tau_1)\},
\label{CCFM-Sudakov-sing}
\end{split}
\end{equation}
where
\begin{equation}
\begin{split}
 \varrho(\tau) 
    &=\; e^\tau (\tau-\tau_0-1)+e^{\tau_0}
\\  &= (t-\ln\Lambda_0)\{\ln((t-\ln\Lambda_0)/(t_0-\ln\Lambda_0))-1\}
     +(t_0-\ln\Lambda_0),
\end{split}
\end{equation}
$\varrho(\tau_0)\equiv0$ and
we have used as previously $t=t(q)=\ln q$, $\tau(t)=\ln(t-\ln\Lambda_0)$,
while $\tau_0=\tau(t_0)=\tau(t(q_0))$ is related to $q_0$ in the IR cut
$1-z_i<\epsilon(q_i,q_0)=q_0/q_i$.
The above is the part (dominant one) of the virtual form factor, but it
suggests clearly another change of the evolution variable $\tau\to\varrho$
at the MC generation level of {\em primary} distributions,
before applying the MC weight,
\begin{equation}
  \varrho(\tau(q)) = \varrho(q;q_0,\Lambda_0) 
   = e^{\tau(q)} \left[\tau(q)-\tau(q_0)-1\right] +e^{\tau(q_0)}\,,
\end{equation}
which could increase the MC efficiency due to
\begin{equation}
  d\varrho = d\tau\;e^\tau (\tau-\tau_0)
 = dt\; 
  \int_0^1 dz\; \frac{\alpha_s^{(0)}((1-z)q)}{\alpha_s^{(0)}(t_A)}\; 
  \frac{\Theta(1-z-q_0/q)}{1-z}     
\end{equation}

The function $\varrho(t)$ cannot be inverted analytically.
However, we have implemented in the MC a simple and fast
subprogram for the numerical inversion $t(\rho)=\varrho^{-1}(t)$.
Thus, the $\varrho$-function can be used to generate
the variable $t$ for every MC event.
The efficiency of the MC is improved
even more by performing importance sampling for the
variable $z$ as well.
To this end we calculate analytically the integral:
\begin{equation}
  \begin{split}
\phi(t,t_0) &=  \int\limits_0^{1} dz\;
     \frac{\alpha_s^{(0)}(t+\ln(1-z))}{\pi} \,
     \frac{\Theta(1-z-\epsilon(t))}{1-z}
\\&=\frac{2}{\beta_0}
     \int\limits^0_{t_0-t} d \ln x\;
     \frac{1}{t-\ln\Lambda_0 +\ln x}
   =\frac{2}{\beta_0}\;
   \ln(\ln x + t -\ln\Lambda_0) \bigg|_{\ln x =t_0 -t}^{\ln x = 0}
\\&=\frac{2}{\beta_0}\;\ln\frac{t -\ln\Lambda_0}{t_0 -\ln\Lambda_0} 
   =\frac{2}{\beta_0}\;\left[\zeta(z=0) - \zeta(z=1-e^{t_0-t})\right],    
\end{split}
\end{equation}
where 
\begin{equation}
\zeta(z) = \ln\left(\,\ln(1-z) + t - \ln\Lambda_0\right).
\end{equation}
The function $\zeta(z)$ can be inverted analytically, giving rise to
\begin{equation}
z(\zeta) = \zeta^{-1}(z) = 1 - \exp(\,e^{\zeta} - t + \ln\Lambda_0).
\end{equation}
Therefore, the variable $\zeta$ can be used to generate $z$ 
according to the {\em primary} MC distribution.

In order to check how this works in practice, we started from
pure gluon-strahlung in the LL approximation. 
From the gluon--gluon momentum kernel we retained only 
the part of $z\Peu$ singular in $(1-z)$, which we denote as 
$z\Peu^{A\Theta}_{GG}(z,t;t_0)$, 
where
\begin{equation}
  \begin{split}
  \Peu^{A\Theta}_{GG}(z,t;t_0)  
  & = \frac{\alpha_s^{(0)}(t+\ln(1-z))}{\pi}\,A_{GG}^{(0)}
  \left[\frac{1}{1-z} + \frac{1}{z}\right]\,\Theta(1-z-\epsilon(t)).
\label{peu}
\end{split}
\end{equation}
Note, that eq.~(\ref{peu}) is given for $\Peu$, so also the $1/z$
singularity is present%
\footnote{
  This splitting function is just the kernel of the one-loop CCFM equation
  as formulated in Ref.~\cite{Marchesini:1991zy}. The only difference is that
  in Ref.~\cite{Marchesini:1991zy} the terms $1/(1-z)$ and $1/z$ are multiplied
  by the running QCD coupling $\alpha_s$ with different arguments, while
  in the above equation we use the same argument of $\alpha_s$ for both
  terms. Such a choice is suggested by the NLL corrections within BFKL scheme,
  see e.g.~\cite{Jung:2000hk}.
}.
Let us remark that the gluon momentum sum rule
\begin{equation}
\int_0^1 dz\, {z\Peu}^{A}_{GG}(z,t;t_0) = 0
\label{gluon-sumrule}
\end{equation}
is fulfilled by the form factor of eq.~(\ref{CCFM-Sudakov-sing}).
As a result, the MC weight from this part of the algorithm is exactly equal 
to $1$.
In fact, the total event weight may slightly differ from $1$, if
generation of $x_0$ and $K_0$
according the initial distribution $D_{K_0}(q_0,x_0)$
is done using {\tt FOAM}~\cite{foam:2002}
in the mode of weighted MC events.

\subsection{Non-singular terms in the gluon kernel}

In the previous subsection we have considered a simplified case of
gluon-strahlung, retaining only the singular terms of the gluon kernel.
Now we are going to extend this analysis by including also
the non-singular terms.
It will be demonstrated on the example of the $GG$ kernel.
The case of the $QQ$ kernel can be treated in the same way. 
The full LL gluon--gluon kernel corresponding
to the real-gluon emission with $z$-dependent $\alpha_s$ reads   
\begin{equation}
\Peu^{\Theta}_{GG}(z,t;t_0) = \frac{\alpha_s(t,z)}{\pi}\, 2C_A
\left[\frac{1}{1-z} + \frac{1}{z} - 2 + z(1 - z) \right]
\Theta(1 - z - \epsilon(t)).
\end{equation}
For the kernel corresponding to the gluon momentum distribution we get
\begin{equation}
\begin{split}
z\Peu^\Theta_{GG}(z,t;t_0) 
& = \frac{\alpha_s(t+\ln (1-z))}{\pi}\, 2C_A \,
\left[\frac{1}{1-z} + \tilde{F}_{GG}(z) \right]\, \Theta(1 - z - \epsilon(t)), 
\\ 
\tilde{F}_{GG}(z) & =  z[-2 + z(1 - z)].
\end{split}
\end{equation}
Enforcing the validity of the gluon momentum sum rule of 
Eq.~(\ref{gluon-sumrule}), the following Sudakov
form-factor is obtained
\begin{equation}
 \Phi_{G}^F(t_2,t_1;t_0) 
= \int_{t_1}^{t_2} dt \int_0^1 dz \,z\Peu^\Theta_{GG}(z,t;t_0).
\end{equation}
For the non-singular part of the gluon kernel, however, the integral over $z$ 
cannot be calculated analytically. After integrating over $t$, for the
non-singular part of the Sudakov form-factor exponent $\Phi_G$, 
labeled as $\Phi_G^{F}$, we obtain easily
\begin{equation}
\begin{split}
 \Phi_G^{F}(t_2,t_1;t_0) =
\frac{2}{\beta_0}\, 2C_A\, 
& 
\bigg\{ \int_0^{1-\exp(t_0-t_1)} dz 
\ln \left(\frac{t_2 -\ln\Lambda_0 + \ln(1-z)}{t_1 -\ln\Lambda_0 + \ln(1-z)}
\right)\,
\tilde{F}_{GG}(z)
\\ 
& + 
\int_{1-\exp(t_0-t_1)}^{1-\exp(t_0-t_2)} dz 
\ln\left(\frac{t_2 -\ln\Lambda_0 + \ln(1-z)}{t_0 -\ln\Lambda_0}\right) \,
\tilde{F}_{GG}(z) \bigg\}
\\
=\frac{2}{\beta_0}\, 2C_A\, 
& 
\bigg\{ \int_0^{t_1-t_0} du\, e^{-u}\, 
\ln \left(\frac{t_2 -\ln\Lambda_0 - u}{t_1 -\ln\Lambda_0 - u}\right)\,
\tilde{F}_{GG}(1-e^{-u})
\\ 
& +  
\int_{t_1-t_0}^{t_2-t_0} du \, e^{-u}\,
\ln \left(\frac{t_2 -\ln\Lambda_0 - u}{t_0 -\ln\Lambda_0}\right) \,
\tilde{F}_{GG}(1-e^{-u}) \bigg\}, 
\end{split}
\end{equation}
where $u = - \ln (1-z)$.
Numerical evaluation of
the above one-dimensional integral can be done
quite precisely ($\sim 0.01\%$) and rather quickly,
as compared to the time of generating a single MC event,
particularly in its latter form.
The singular part, $\Phi_G^{A}$, of the Sudakov form-factor exponent is given
by Eq.~(\ref{CCFM-Sudakov-sing}).
The total Sudakov form-factor exponent is now
\begin{equation}
\Phi_G = \Phi_G^{A} + \Phi_G^{F}\,.
\end{equation}

The non-singular gluon kernel terms can be easily implemented in the forward 
Markovian algorithm of the previous subsection through appropriate MC weights.
For the real-gluon radiation the corresponding weight is
\begin{equation}
W_z =  1 - z(1-z)\,[\,2 - z(1 - z)\,],
\end{equation}
This weight is very well-behaved: $\frac{7}{16} \leq W_z \leq 1$. 
The virtual-gluon contribution has to be compensated with the weight
\begin{equation}
W_\Delta =\exp(\bar{\Delta}_G),\;\;\;\;
\bar{\Delta}_G = {\Phi}^A_G - \Phi_G = - \Phi_G^{F}\,.
\end{equation}

\subsection{Quark-gluon transitions}

Although the original CCFM equation was formulated for gluons only,
we may try to extend it to quarks and allow for quark--gluon
transitions.
The treatment of the quark kernels is identical to the case of 
gluon kernels described in the
previous subsections. As for the case of $QG$ transitions,
we intend to apply the same importance sampling for generation of 
the $t$ and $z$ variables as described in Subsection~\ref{SubSec:step2}, 
using at the primary MC generation level the following 
approximate kernels
\begin{equation}
z\bar{\Peu}^\Theta_{IK}(z,t;t_0)
= \frac{\alpha_s^{(0)}(t+\ln(1-z))}{\pi}
     \; \frac{\Theta(1-z-\epsilon(t))}{1-z}\; 
      \left[ \delta_{IK} A^{(0)}_{KK} +
       \bar{F}_{IK}^{(0)}
       \right]\,,
\end{equation} 
where $\bar{F}_{IK}^{(0)} \equiv \max_z F_{IK}^{(0)}(z)$ 
are given in Appendix~C of Ref.~\cite{Golec-Biernat:2006xw}.
This means that at the low MC level we artificially include singular
factors $1/(1-z)$ 
for non-diagonal transitions that are not present in the corresponding 
exact kernels. The above approximation is then compensated by the 
MC weight being the ratio of the exact to approximate kernels
\begin{equation}
w_{IK}^{\Pcal} = \frac{\Peu^\Theta_{IK}(z,t;t_0)}
                      {\bar{\Peu}^\Theta_{IK}(z,t;t_0)}\:\leq 1\,.
\end{equation} 
The loss of efficiency due to this artificial modification is rather small,
whereas the gain in simplicity of the
algorithm is significant.

For the exponent of the Sudakov form-factor we get
\begin{equation}
\begin{split}
 \Phi_K(t_2,t_1;t_0)= 
\frac{2}{\beta_0} \,
\Bigg\{ 
& 
A_{KK}^{(0)}\;\; \left[\varrho(t_2)-\varrho(t_1)\right]
\\
&+ 
 \int_0^{t_1-t_0} du\, e^{-u}\, 
\ln \left(\frac{t_2 -\ln\Lambda_0 - u}{t_1 -\ln\Lambda_0 - u}\right)\,
\sum_J F_{JK}^{(0)}(1-e^{-u})
\\ 
& +  
\int_{t_1-t_0}^{t_2-t_0} du \, e^{-u}\,
\ln \left(\frac{t_2 -\ln\Lambda_0 - u}{t_0 -\ln\Lambda_0}\right) \,
\sum_J F_{JK}^{(0)}(1-e^{-u})\Bigg\}\,, 
\end{split}
\end{equation}
where the functions $F_{JK}^{(0)}(z)$ are given Appendix~C 
of Ref.~\cite{Golec-Biernat:2006xw}. Again, the non-singular terms in the
above form factor have to be integrated numerically.

\subsection{Numerical tests}

We have performed comparisons of the MC solution of the above
evolution equations, implemented the program {\tt EvolFMC}~\cite{EvolFMC:2006},
with the solution provided by the non-MC program 
{\tt APCheb40}~\cite{APCheb40} in which the same $z$-dependent strong
coupling has also been implemented.
In both cases we have evolved the singlet PDFs for gluons
and three doublets of massless quarks from $Q_0=1\,$GeV
to $Q=10,100,1000\,$GeV.
We have used the following parameterization of the starting 
parton distributions in the proton at $Q_0=1\,$GeV:
\begin{equation}
\label{startev}
  \begin{split}
    xD_G(x)        &= 1.9083594473\cdot x^{-0.2}(1-x)^{5.0},\\
    xD_q(x)        &= 0.5\cdot xD_{\rm sea}(x) +xD_{2u}(x),\\
    xD_{\bar q}(x) &= 0.5\cdot xD_{\rm sea}(x) +xD_{d}(x),\\
    xD_{\rm sea}(x)&= 0.6733449216\cdot x^{-0.2}(1-x)^{7.0},\\
    xD_{2u}(x)     &= 2.1875000000\cdot x^{ 0.5}(1-x)^{3.0},\\    
    xD_{d}(x)      &= 1.2304687500\cdot x^{ 0.5}(1-x)^{4.0},
  \end{split}
\end{equation}

\begin{figure}[!ht]
  \centering
  {\epsfig{file=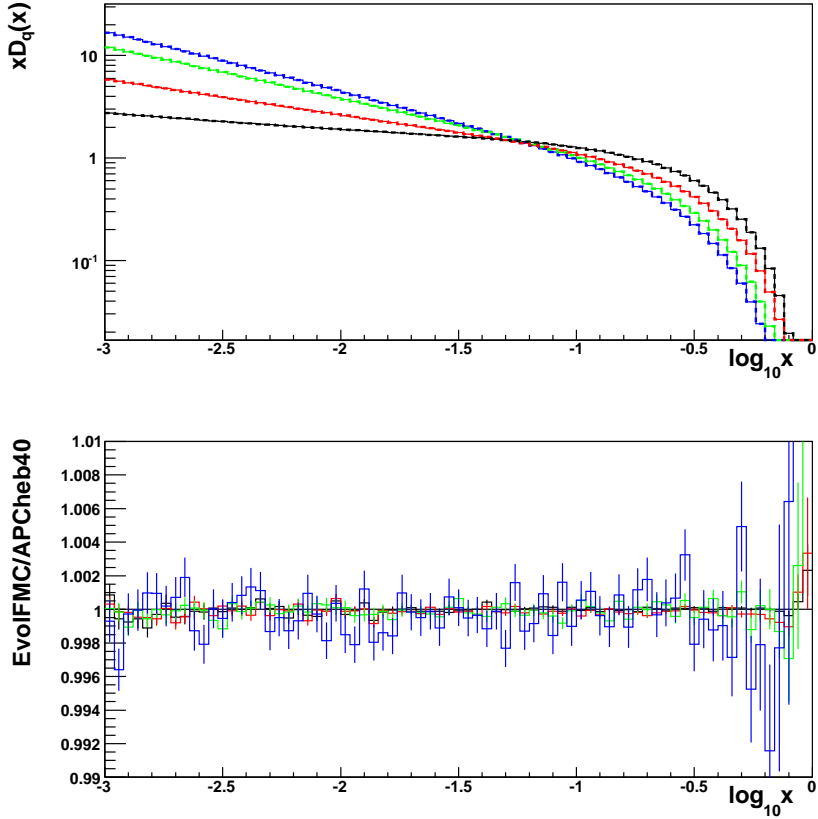, width=120mm}}
  \caption{\sf
    The upper plot shows the singlet quark distribution $xD_{q}(x,Q)$ 
    evolved from $Q_0=1\,$GeV (black) to $Q=10$ (red), 
    $100$ (green) and $1000$ (blue) GeV,
    obtained from {\tt EvolFMC} (solid lines) and  
    {\tt APCheb40} (dashed lines, hardly distinguishable), 
    while the lower plot shows their ratio.
  }
  \label{fig:MMC-APCheb40_quarks}
\end{figure}

\begin{figure}[!ht]
  \centering
  {\epsfig{file=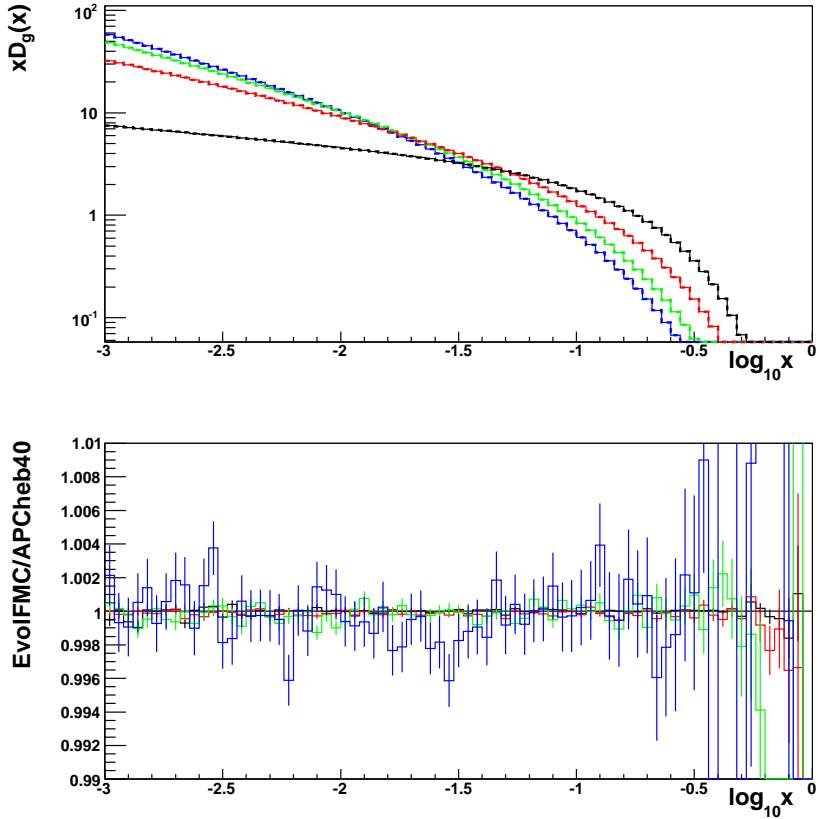, width=120mm}}
  \caption{\sf
    The upper plot shows the gluon distribution $xD_{q}(x,Q)$ 
    evolved from $Q_0=1\,$GeV (black) to $Q=10$ (red), 
    $100$ (green) and $1000$ (blue) GeV,
    obtained from {\tt EvolFMC} (solid lines) and  
    {\tt APCheb40} (dashed lines, hardly distinguishable), 
    while the lower plot shows their ratio.
  }
  \label{fig:MMC-APCheb40_gluons}
\end{figure}

In Fig.~\ref{fig:MMC-APCheb40_quarks} we show the resulting quark
distributions obtained from the two programs (the upper plot)
as well as their ratios (the lower plot).
As one can see, these two calculations agree at the level of $0.2\%$,
except for the $x$-region very close to $1$, where the MC statistics is low.
The similar agreement has been found for the resulting gluon distributions, 
shown in Fig.~\ref{fig:MMC-APCheb40_gluons}. 
The precision of the presented results is limited by the statistical errors
of the MC calculations.

\section{One-loop CCFM equations of Kwieci\'nski}

The CCFM equations \cite{CCFM} describe the evolution of an unintegrated 
parton distributions $d_K(x,k^T,Q)$ which depend on the parton transverse
momentum $k^T$ in addition to the longitudinal momentum fraction
$x$ and the scale $Q$. They are related to the integrated
parton distribution of the DGLAP equations $D_K(x,Q)$ through
the following relation
\begin{equation}
x D_K(x,Q) = \int d^2 k^T\, x d(x,k^T,Q) 
\end{equation}

Originally, the CCFM equations were derived for the unintegrated gluon
distribution. In the all-loop approximation,
they also take  into account angular ordering (coherence)
in the initial state gluon cascade for both large and small
values of $x$. This leads to a
new non-Sudakov form factor that sums virtual corrections for small $x$. 

In the one-loop approximation, the angular ordering at small $x$ and the 
corresponding virtual corrections are subleading. Thus, the resulting 
equations contain
only coherence in the real parton emissions for large $x$,
including only the Sudakov form factor which  resumes virtual corrections.
At small $x$ the standard DGLAP transverse momentum ordering appears.
The one-loop   CCFM equations for both quark and 
gluon unintegrated distributions,  written for the first time 
in \cite{Kwiecinski:2002}, take the following form
\begin{align}
x d_{K}(x,k^T,Q) &= xd^0_{K}(x,k^T)
\notag
\\ 
& +\int\frac{d^2\vec{q}}{2\pi q^2}
\theta(q-q_{0})\theta(Q-q)
\int\limits_x^1\frac{dz}{z}\, z\Peu_{KJ}(q,z) 
   \,\frac{x}{z}\,d_{J}\Bigl(\frac{x}{z},l^T,q\Bigr),
\notag
\\
\label{kwiecinski}
\vec{l}^T &= \vec{k}^T+ (1-z)\vec{q}\,.
\end{align}
This equation is slightly more general than the original equation from
Ref.~\cite{Kwiecinski:2002} since it allows for the coupling constant
(hidden in the splitting functions $\Peu_{KJ}(q,z)$)
which depends both on $z$ and $q$. 
The iterative solution to eq.~(\ref{kwiecinski}) reads
\begin{align}
\notag
x d_{K}&(x,k^T,Q) = xd^0_{K}(x,k^T)
+\sum_{n=1}
  \int\limits_0^1 dz_0
  \int d^2\vec{k}^{T}_{0}
   \sum_{K_0,\ldots,K_{n-1}}
\\
\notag
&\Biggl[
\prod_{i=1}^n
\int\limits_{q_0}^Q\frac{d^2\vec{q_i}}{2\pi q_i^2}\,
\theta(q_i-q_{i-1}) 
\int\limits_0^1{dz_i}\, z_i\Peu_{K_i K_{i-1}}(q_i,z_i) 
\Biggr]
   xd^0_{K_0}\Bigl(z_0,{k}^{T}_{0}\Bigr)
\\&
\delta\Bigl(x-\prod_{i=0}^nz_i\Bigr)\,
\delta^2\Bigl(
       \vec{k}^{T}-\vec{k}^{T}_{0} +\sum_{i=1}^n(1-z_i)\vec{q}_{i}\Bigr)\,,
\label{kwiecinski-iter}
\end{align}
where $K\equiv K_n$.
As compared to the basic iterative solution of the DGLAP-type equation,
eq.~(\ref{eq:Iter6}), the above series differs only by the presence
of the independent angular integrals in 
$d^2 q_i$ and by the presence of the delta function 
$\delta^2(\vec{k}^{T}-\dots)$,
which in the Markovian-type algorithm plays only a ``spectator'' role.
In addition, in eq.~(\ref{kwiecinski-iter}) the Sudakov form factor
has not been explicitly worked out, as in eq.\
(\ref{eq:Iter6}), and the initial condition
$xd^0_{K}$ is now
${k^T}$-dependent. Having that in mind, we reorganize
eq.~(\ref{kwiecinski-iter})  in the
form similar to the DGLAP solution, eq.~(\ref{eq:Iter6}),
\begin{equation}
  \label{eq:Iter:phi}
  \begin{split}
  x d_{K}&(x,k^T,t)= e^{-\Phi_K(t,t_0)} xd^0_{K}(x,k^T,t_0)
  +\sum_{n=1}^\infty \;
  \int\limits_0^1 dx_0\;
  \int d^2\vec{k}^{T}_{0}
   \sum_{K_0,\ldots,K_{n-1}}
\\&
      \prod_{i=1}^n \bigg[ \int\limits_{t_0}^t dt_i
    \; \Theta(t_i-t_{i-1}) \int \frac{d\phi_i}{2\pi} \int\limits_0^1 dz_i\bigg]
      e^{-\Phi_K(t,t_n)}
      \prod_{i=1}^n 
          \bigg[ 
                 z_i\Peu_{K_i K_{i-1}}^\Theta (t_i,z_i) 
                 e^{-\Phi_{K_{i-1}}(t_i,t_{i-1})} \bigg]
\\&~~~\times
      xd^0_{K_0}\Bigl(z_0,{k}^{T}_{0},t_0\Bigr) \,
      \delta\big(x- x_0\prod_{i=1}^n z_i \big)\,
\delta^2\Bigl(
       \vec{k}^{T}-\vec{k}^{T}_{0} +\sum_{i=1}^n(1-z_i)\vec{q}_{i}\Bigr)\,,
  \end{split}
\end{equation}
where as before $q_i=\exp(t_i)$ and $\phi_i$ is the azimuthal angle of
$\vec{q}_i$.  Notice that $t=\ln Q$ plays now the role of the evolution
variable.

It is now transparent that as described in Ref.~\cite{Golec-Biernat:2006xw} 
the forward Markovian algorithm {\tt EvolFMC} can easily be  extended to  
embed the generation of parton transverse momentum $k^T$. 
After $n$ steps of the forward Markovian
evolution the transverse momentum of the {\em off-shell} parton
entering the hard process
becomes
\begin{equation}
\vec{k}^{T}_{n} = \vec{k}^{T}_{0} - \sum_{i=1}^{n} (1-z_i)\vec{q}_i\,,
\end{equation}
where $\vec{k}^{T}_{0}$ is an intrinsic parton transverse momentum and
$\vec{q}_i$ a 2-dimensional evolution variable. The physical
transverse momenta of emitted particles are
\begin{equation} 
(1-z_i)\vec{q}_i\,.
\end{equation} 
In the MC evolution $\vec{q}$ is constructed as follows:
\begin{equation}
  \vec{q}_i = e^{t_i}\,(\cos\phi_i,\sin\phi_i),
\end{equation}
where $t_i$ is an evolution variable and $\phi_i$ is an azimuthal angle
generated at each evolution step from a flat distribution in the range
$[0,2\pi]$. 

In our numerical tests, which will be reported in the next section,
the intrinsic 
parton transverse 
momentum $k^{T}_{0}$ is generated at the initial evolution scale $t_0$
from the following $x$-independent distribution
\begin{equation}
\label{startkt}
\frac{1}{k_0^2}\, \exp\left\{\frac{-(k^{T}_{0})^2}{k_0^2}\right\},  
\end{equation}
where $k_0$ is some adjustable parameter, set to $1\,{\rm GeV}$  
in our tests.
 
\section{Numerical results for $k^T$-distributions}

In the following we present results for the unintegrated parton
distributions 
as functions of the transverse momentum $k^T$, generated by the forward Markovian MC
program {\tt EvolFMC}~\cite{EvolFMC:2006} according to the one-loop CCFM 
equation described in the previous section. 
The starting point of the evolution
is $Q_0 = 1\,$GeV and the initial conditions are specified in
eqs.~(\ref{startev}) and (\ref{startkt}).

\begin{figure}[!ht]
  \centering
 {\epsfig{file=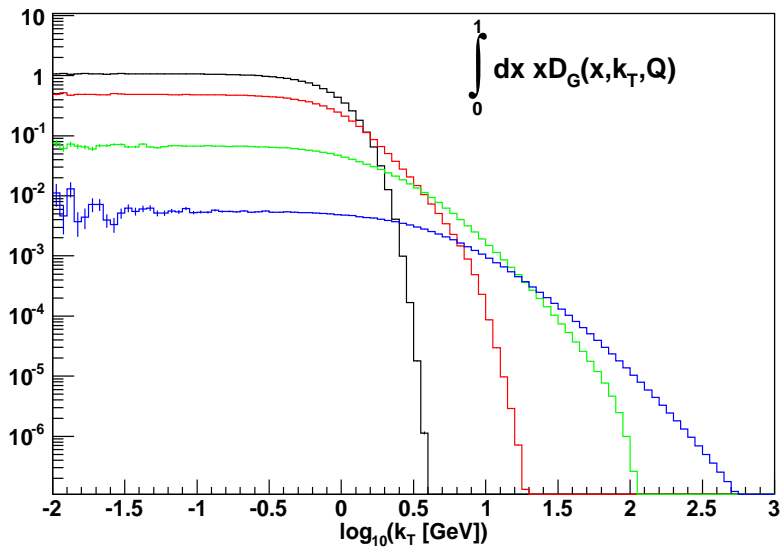, width=120mm, height=80mm}}
  \caption{\sf
    Gluon $k^T$ distributions integrated over $x$ for the one-loop CCFM 
    equation of Marchesini--Webber, obtained from {\tt EvolFMC} 
    for Q = 1 (black), 10 (red), 100 (green) and 1000 (blue) GeV.
    }
  \label{fig:g-logkt-xint-mw}
\end{figure}

\begin{figure}[!ht]
  \centering
   {\epsfig{file=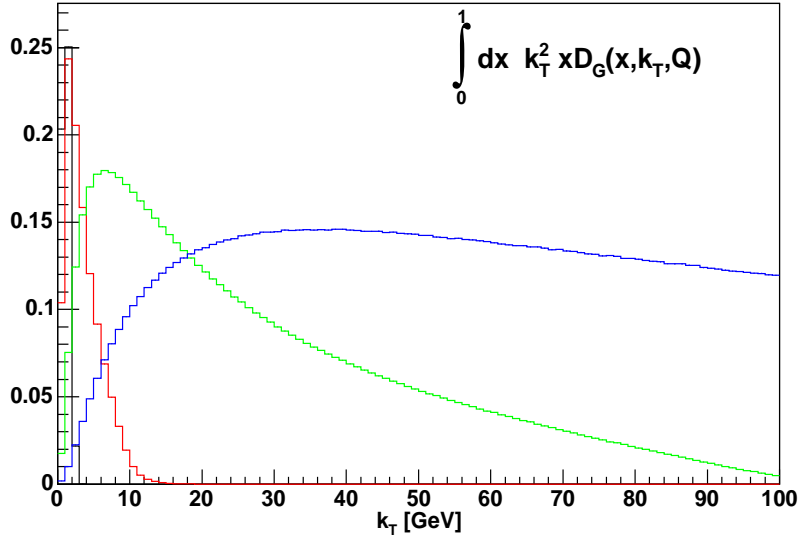, width=120mm, height=80mm}}
  \caption{\sf
    Gluon $k^T$ distributions multiplied by $(k^T)^2$ integrated over $x$ 
    for the one-loop CCFM equation of Marchesini--Webber, 
    obtained from EvolFMC for Q = 1 (black), 
    10 (red), 100 (green) and 1000 (blue) GeV.
    }
  \label{fig:g-kt-xint-mw}
\end{figure}

\begin{figure}[!ht]
  \centering
  {\epsfig{file=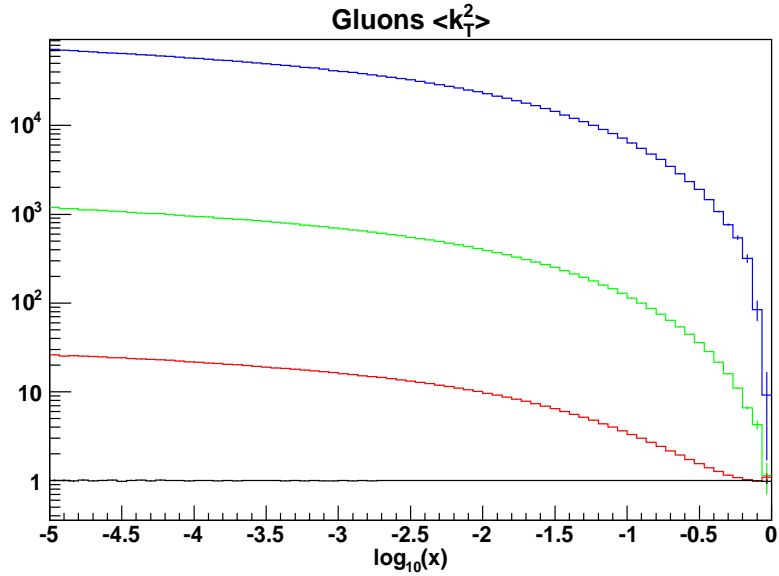, width=120mm, height=80mm}}
  \caption{\sf
    The average  $(k^T)^2$ of gluon as a function of $x$ for the one-loop CCFM 
    equation of Marchesini--Webber, obtained from {\tt EvolFMC} 
    for Q = 1 (black), 10 (red), 100 (green) and 1000 (blue) GeV.
    }
  \label{fig:g-kt2ave-logx-mw}
\end{figure}

In Figs.~\ref{fig:g-logkt-xint-mw} and \ref{fig:g-kt-xint-mw} we
show the gluon distribution obtained from the CCFM equation with
gluons only, as given by  Marchesini and Webber in
Ref.~\cite{Marchesini:1991zy}. The distribution was additionally integrated
over $x$. For rising values of the scale $Q$ such a gluon distribution
moves towards large values of $k^T$,   becoming at the same time less
sensitive to the low $k^T$ region. This observation is summarized in 
Fig.~\ref{fig:g-kt2ave-logx-mw} where we present the average gluon $(k^T)^2$ 
as a function of $x$ for $Q = 1,\, 10,\, 100$ and $1000\,$GeV.

%

\begin{figure}[!ht]
  \centering
  {\epsfig{file=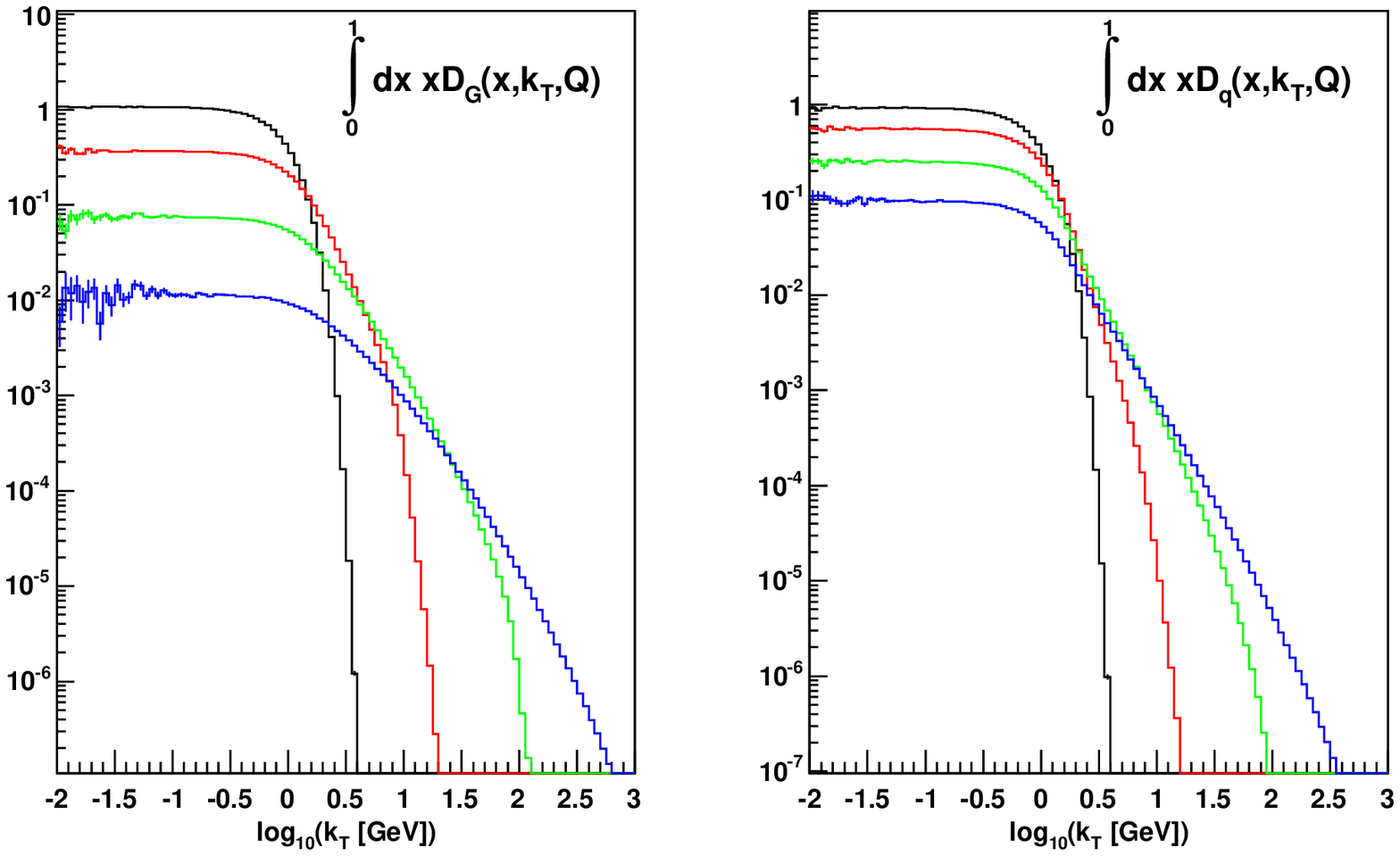, width=150mm, height=80mm}}
  \caption{\sf
    Gluons and quarks $k^T$ distributions integrated over $x$ for the one-loop 
    CCFM equation of Kwieci\'nski, obtained from {\tt EvolFMC} 
    for Q = 1 (black), 10 (red), 100 (green) and 1000 (blue) GeV.
    }
  \label{fig:gq-logkt-xint-k}
\end{figure}

\begin{figure}[!ht]
  \centering
  {\epsfig{file=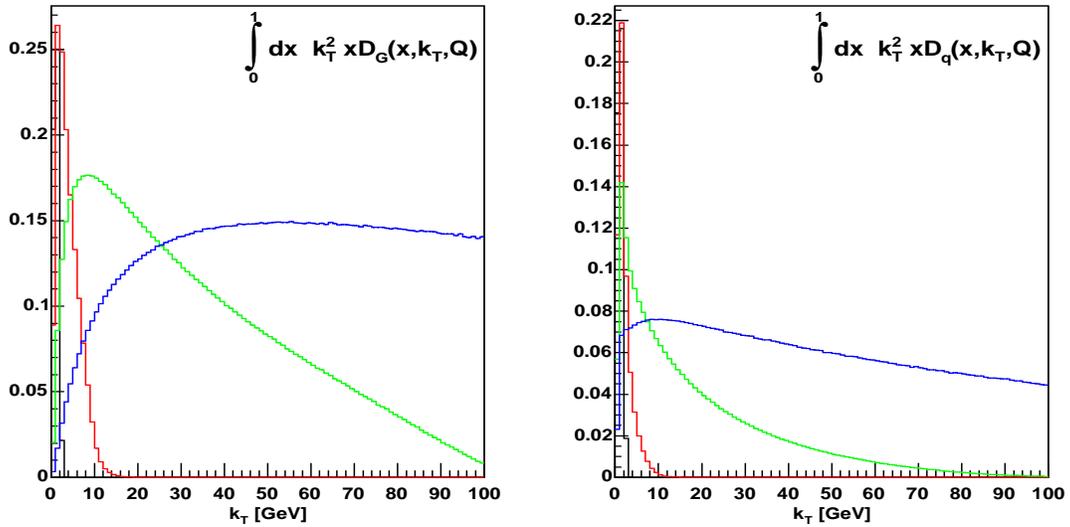, width=150mm, height=80mm}}
  \caption{\sf
    Gluons and quarks $k^T$ distributions multiplied by $(k^T)^2$ integrated 
    over $x$ for the one-loop CCFM equation of Kwieci\'nski, 
    obtained from {\tt EvolFMC} for Q = 1 (black), 
    10 (red), 100 (green) and 1000 (blue) GeV.
    }
  \label{fig:gq-kt-xint-k}
\end{figure}

\begin{figure}[!ht]
  \centering
 {\epsfig{file=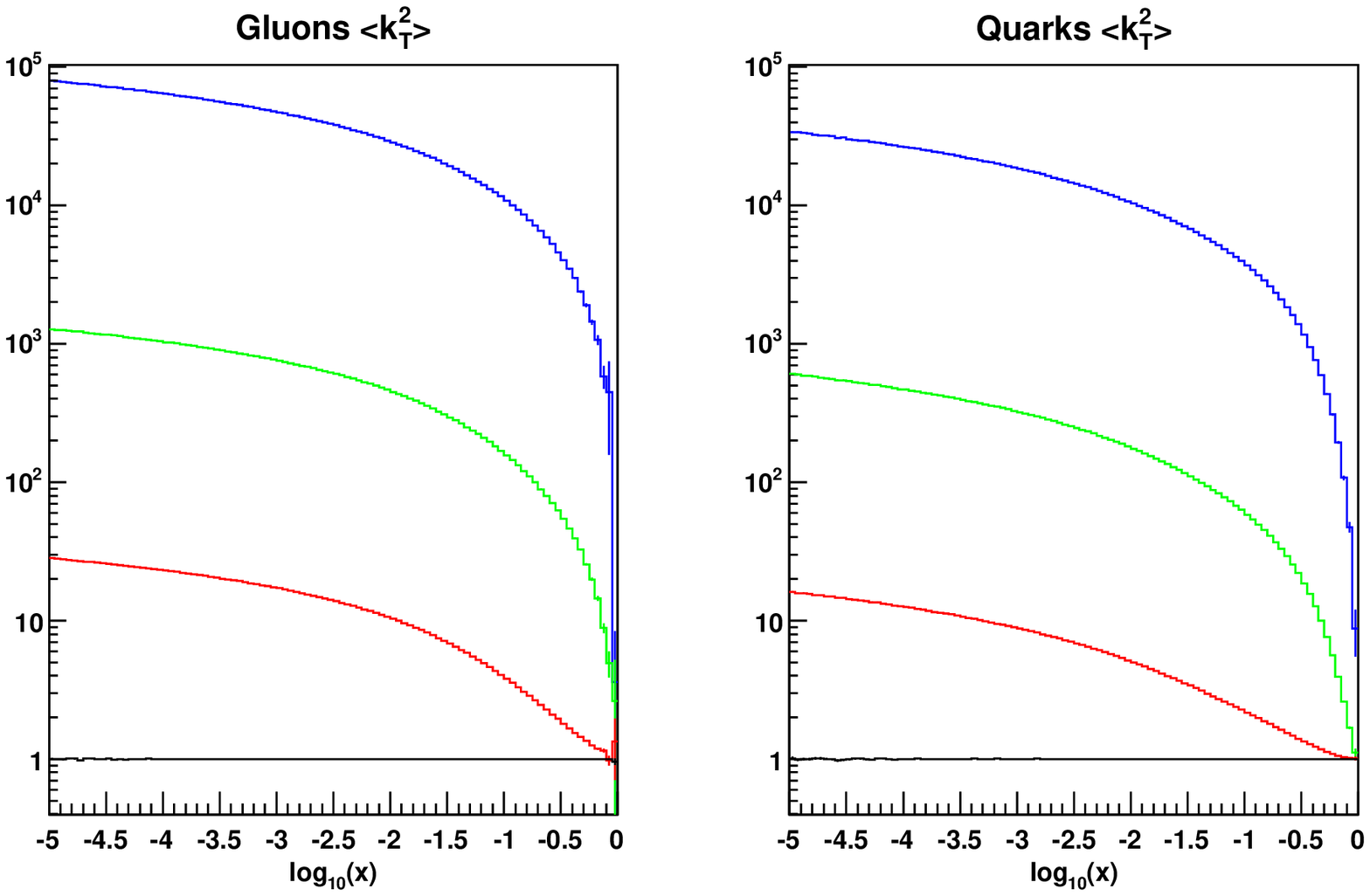, width=150mm,  height=85mm}}
  \caption{\sf
    The average $(k^T)^2$ of gluons and quarks as a function of $x$ for the 
    one-loop CCFM 
    equation of  Kwieci\'nski, obtained from {\tt EvolFMC} for Q = 1 (black), 
    10 (red), 100 (green) and 1000 (blue) GeV.
    }
  \label{fig:gq-kt2ave-logx-k}
\end{figure}

In Figs.~\ref{fig:gq-logkt-xint-k} and \ref{fig:gq-kt-xint-k}, 
we show the results of the same studies for the
full one-loop CCFM equation for the singlet quark and gluon distributions, given by
Kwieci\'nski {\it et al.} in Ref.~\cite{Kwiecinski:2002}. The change of the
distributions (integrated over $x$) with the hard scale $Q$ is the
same as in the previous case. Both the quark and gluon distributions
become less sensitive to the ``soft'' $k^T$ values  moving towards
the region of large   $k^T$'s. This is also visible in Fig.~\ref{fig:gq-kt2ave-logx-k}
for the average transverse momentum for gluons and quarks. As
expected, the average $k^T$ rises when $x\to 0$ \cite{Kwiecinski:2002}.


\section{Summary and outlook}

In this work we have presented a systematic extension of the Monte
Carlo algorithm that solves the DGLAP equation into the algorithm solving
the one-loop CCFM equation.
To this end two technical problems have been solved:
the coupling constant has become $z$-dependent and the evolution
in transverse momenta has been added, in addition
to the evolution in the longitudinal momentum fractions $x$ in the parton distributions.
The modification of the coupling constant has lead to the need of solving
numerically the transcendental equation.
First numerical results have been presented, confirming the known
observations on CCFM unintegrated PDFs.

The presented algorithm for the one-loop CCFM evolution we consider as the
first step in extending the other types of the QCD evolution equations
available in our MC programs beyond that of the DGLAP type.
In particular, implementing the complete CCFM evolution,
both in the Markovian and non-Markovian (constrained) algorithms,
see ref.~\cite{Jadach:2007singleCMC}, is now in an advanced stage.
The presently implemented one-loop CCFM option
will be used in the forthcoming studies
of various aspects of the QCD evolution equations
according to several different schemes.

\vspace{10mm}
\noindent
{\Large\bf Acknowledgments}\\
We would like to thank Z.\ W\c{a}s for the useful discussions.
We thank for warm hospitality of the CERN PH Department were part
of this work was done. The partial financial support of the 
 MEiN research grant~1~P03B~028~28 (2005-08) is acknowledged.


\end{document}